\documentclass[11pt]{article}

\usepackage{lineno,hyperref}
\usepackage{times}

\usepackage{graphicx,xcolor}
\usepackage{amsfonts}
\usepackage{latexsym}
\usepackage{amsmath}
\usepackage{bm}
\usepackage[utf8]{inputenc} 

\newcommand{\rmd}{\mathrm{d}}
\newcommand{\rmj}{\mathrm{j}}
\newcommand{\rme}{\mathrm{e}}
\newcommand{\rmi}{\mathrm{i}}
\newcommand{\ansatz}{\textit{ansatz }}

\begin{document} 

\begin{titlepage}


\vspace{10pt}

\begin{center}

{\Large\bf
Fundamental physics tests using the propagation of GNSS signals
}


\vspace{50pt}

Bruno Bertrand\footnote{\em bruno.bertrand@oma.be} and 
Pascale Defraigne

\vspace{20pt}

Royal Observatory of Belgium (ROB)\\
3, Avenue Circulaire, 1180 Brussels, Belgium

\vspace{60pt}





\begin{abstract}
This paper introduces new tests of fundamental physics by means of the analysis of disturbances on the GNSS signal propagation. We show how the GNSS signals are sensitive to a space variation of the fine structure constant $\alpha$ in a generic framework of effective scalar field theories beyond the Standard Model. This effective variation may originate from the crossing of the RF signals with dark matter clumps and/or solitonic structures. At the macroscopic scale, the subsequent disturbances are equivalent to those which occur during the propagation in an inhomogeneous medium. We thus propose an interpretation of the ``measure'' of the vacuum permeability as a test of fundamental physics. We show the relevance of our approach by a first quantification of the expected signature in a simple model of a variation of $\alpha$ according to a planar geometry. We use a test-bed model of domain walls for that purpose and focus on the measurable time delay in the GNSS signal carrier.
\end{abstract}

\end{center}

\end{titlepage}




\setcounter{footnote}{0}

\section*{Introduction and overview}

It can be inferred from cosmological observations that 23\% of the mass-energy budget of the Universe consists of dark matter (DM) while our ordinary baryonic matter is reduced to 4\% \cite{Aghanim:2018eyx}. Hence a plethora of models have been developed this last decade to account for the microscopic nature of DM. 
Within that context, recent investigations suggest that DM takes the form of compact objects like clusters or macroscopic structures with classical properties. We propose to roughly divide into three categories these dark compact objects: self-interacting solitons, dark clusters and stars, and finally primordial black hole (not described here). Self-interacting solitons can be distinguished into topological and non-topological solitons.

\paragraph{Topological defects (Topological solitons).}
Standard cosmological models predict that macroscopic structures called topological defects could appear during phase transitions in the early ages of the universe, see \cite{Vilenkin:2000jqa} for a review. In particular, domain walls are a special case of topological defects with a planar configuration. 
%
If networks of domain walls exist, they are not necessarily responsible for dark matter \cite{Roberts:2017hla} and they would make only a modest contribution for the dark energy as shown by \cite{Sousa:2015cqa}. However, topological defects, with their associated symmetry breaking, arise very naturally in models of new physics. Therefore they represent an ideal ``test-bed laboratory'' in a phenomenological approach.   
%
%
For example, the symmetron model introduced in \cite{Hinterbichler:2010es}, deals with light scalar fields and admits domain wall solutions not related to the primordial area of the universe, as shown by \cite{Silva:2013sla} in the case of a non trivial coupling to electromagnetism.
%


\paragraph{Non topological solitons.} Non topological solitons are classical solutions which naturally arise in extensions of the Standard Model. These compact macroscopic objects are also running as DM candidates, as shown by \cite{Kusenko:2001vu} for $Q$-balls.

\paragraph{Dark clusters and stars.}
Some theoretical models of cosmic evolution make possible the organization of DM into compact micro-halos \cite{Dror:2017gjq} or clusters \cite{Visinelli:2018wza} whose mass can be as low as $10^{12}$ kg. In other models, gravitationally bound systems of DM can also condensate into soliton-like configurations like axion stars, boson stars or relaxation stars, see \cite{Braaten:2018nag,Eby:2015hsq,Banerjee:2019epw} for a review. Such models of dark objects have a classical behaviour. Moreover, if the associated particles exist, they naturally fit well with the cosmological relic abundance of dark matter. 
\newline



In a fascinating perspective, each of these three types of dark objects could regularly cross the Earth for specific configurations of the parameter space, hence their name: transient DM. This possibility was described independently for different models: topological defects \cite{Pospelov:2012mt,Derevianko:2013oaa}, non topological solitons \cite{JacksonKimball:2017qgk,Roberts:2018agv} or gravitationally bound dark particles \cite{Visinelli:2018wza,Banerjee:2019epw}.
In that case, recent works even raise the possibility that these objects may temporarily be trapped by the external gravitational potential of the Earth or the Sun, as proposed in \cite{Banerjee:2019epw, Xu:2008ep}. These objects
would have strongly elliptical orbits and could regularly
get close to the Earth orbit. Such an hypothesis arouses interest for high precision measurements in the close vicinity of the Earth.

Tests of General Relativity received a great interest during these last two decades, since many theories beyond the Standard Models predict a violation of the Local Position Invariance (LPI). The principle of LPI states that the outcome of any local non-gravitational experiment is independent of the space-time position. In the case the DM transients interact (weakly) with electromagnetism (EM), as suggested by numerous effective models, the violation of the LPI should be manifest trough the space variation of the fine structure constant $\alpha$ that governs the strength of electromagnetic interaction. Such variation of fundamental constants should affect the frequency of atomic clocks. Hence, a novel idea which has been recently developed in the literature, for example in \cite{Pospelov:2012mt,Derevianko:2013oaa,Banerjee:2019epw}, is to twist the network of atomic clocks onboard Global Navigation Satellite System (GNSS) satellites and on the Earth and transform this network into DM detectors. This strategy is always winning with a reduced cost. Even though no signature is detected, such an analysis will reduce the actually large parameter space. 
%
However, this method is strongly model-dependent. In general, the detection of such transient signatures by atomic clocks is mainly restricted to soliton-like structures of dark matter. Strong reductions of the parameter space have been recently obtained by \cite{Wcislo:2018ojh,Roberts:2017hla,Roberts:2019sfo} using clock comparison via GNSS and optical frequency transfer techniques for the two latter references. Nevertheless the constraints brought by these analyses deal with the coupling of the dark structures to ordinary matter and not with their abundance. Moreover, the sensitivity of these experiments is reduced to solitons of which the typical dimension is large, at least beyond the kilometer scale. 

This paper addresses the possibility of detecting a space variation of $\alpha$ induced by transient objects like DM clumps and/or solitons using the propagation of electromagnetic (EM) waves, in particular the signals from GNSS satellites. In the future this approach will enable to detect a larger variety of DM transient than techniques based on atomic clocks. The further advantage of using the signals is that the interaction time between the transient and the GNSS facilities is now enlarged to the transit time between the satellite and the receiver. We choose in this paper to apply our method to the simple case of domain walls with a quadratic scalar coupling to EM since it offers (1) an easier data correlation of different GNSS signal perturbations in the network and (2) the exact solution of the basic model of DW to describe the variation of $\alpha$. 

This paper is threefold. First, we propose an innovative way to link interactions violating the Local Position Invariance (LPI) in the Maxwell sector to variables as measured in RF engineering systems. Then, we describe the perturbations induced by hypothetical transient domain walls on the propagation of GNSS signals. Finally we quantify the projected subsequent signature in GNSS measurements.

\section{Description of our phenomenological model}
\subsection{Fundamental theoretical framework}
In this section, we apply our method of testing fundamental physics using the propagation of GNSS signals in the case of an hypothetical scalar field $\varphi(x)$ coupled to electromagnetic fields. Starting from a quite generic model of effective scalar field theories beyond the Standard Model, the Maxwell's Lagrangian density reads:
\begin{equation}\label{eq:Maxwell-phi_Lag}
 \mathcal{L} = -\frac{k_0}{4} \left(1 \pm \frac{\varphi^{\mathrm{n}}(x)}{\Lambda^{\mathrm{n}}} \right)\,  F_{\mu\nu}\, F^{\mu\nu} - A_\mu\, J^\mu + \mathcal{L}_{\varphi}(\varphi,\partial_\mu\varphi) + \mathcal{L}_{\varphi\textrm{int}} + \cdots \, ,
\end{equation}
where $\mu=0,1,2,3$ refers to the flat Minkowski spacetime while $k_{\mathrm{i}}$ and $\Lambda$ are dimensional constants. This kind of non minimal coupling is known to induce a violation of the LPI through the spacetime dependence (and dynamics) of the fine structure constant $\alpha$, see \cite{Uzan:2010pm} for a review. The Faraday tensor is written as $F_{\mu\nu} = \partial_\mu A_\nu - \partial_\nu A_\mu$, where $\partial_\mu$ is the derivative with respect to the vector coordinate $x^\mu$. The time component $A_{\mathrm{0}}$ is proportional to the absolute electric potential while the space components $A_{\mathrm{i}}$ are the magnetic vector potential. The ellipses stand for generic (baryonic or dark) matter fields that generates current $J^\mu$ and which can also couple to $\varphi$ through $\mathcal{L}_{\varphi\textrm{int}}$.
The high energy scale $\Lambda$ accounts for the effective character of the model. Finally, $k_0$ is a purely dimensional constant relating the units of electromagnetism to mechanical units. In contradistinction with \cite{Stadnik:2014cea}, we choose the interaction term to be $U(1)$ gauge invariant. The dynamics of $\varphi(x)$ is described by $\mathcal{L}_{\varphi}$. 
\begin{equation}\label{Lag:Goldstone}
\mathcal{L}_{\varphi} = \frac{1}{2} \partial_\mu \varphi\, \partial^\mu \varphi - V\left(\varphi\right)\, , 
\end{equation}
where $V\left(\varphi\right)$ is the self-interacting potential.


\subsection{Effective Maxwell's equations in a medium}
It was noted long ago that the non-minimal coupling of a scalar field to the Maxwell theory possesses formal analogies with macroscopic Maxwell equations in a material medium (see the review of \cite{Uzan:2010pm} and ref. therein). However this analogy was scarily exploited so far in the literature in terms of propagation of electromagnetic waves (contrary to the Standard Model Extension and the violation of the Local Lorentz Invariance in \cite{Kostelecky:2002hh}). The notable exception was the works of \cite{Stadnik:2014cea} and \cite{Hees:2014lfa} but they restricted their work to the eikonal approximation which hides the optically active effects. Here we show such effects in theories where the coupling of a scalar field to EM violates the LPI. We mainly focus our attention on GNSS signal propagation and hence provide an interpretation in terms of GNSS observable.


``Microscopic'' Maxwell's equations describe how the total charge and current generate respectively the electric $\vec{E}$ and magnetic $\vec{B}$ fields, including the free charge density $\rho_\textrm{f}$ and current density $\vec{J}_\textrm{f}$ screened by the internal charge distribution and movement in a given medium. By contrast, the ``macroscopic'' Maxwell's equations introduce new ancillary fields, that we call in this paper the magnetization field $\vec{H}$ and displacement field $\vec{D}$. So this macroscopic description includes in an effective formulation all the effects of quantum interactions inside materials, giving rise to the empirical influence of bound currents and charges. 
%
Thus, only phenomenological parameters characterize electromagnetic properties of materials (e.g. polarisation, magnetisation). 

By analogy, we will consider effects inherent to a classical configuration of the fields in fundamental physics as the effective phenomenology of a medium. The examples already mentioned cover solitonic structure along with effective descriptions of self-interacting or gravitationally bound condensates of particles. In that way, effective effects inherent to new physics are encoded into the fields $\vec{H}$ and $\vec{D}$ so that we associate a phenomenological effective description to a macroscopic effect likewise it happens in collective phenomena of condensed matter.


Starting from the Lagrangian density (\ref{eq:Maxwell-phi_Lag}), the least action principle can be applied to obtain the (modified) Gauss's and Maxwell-Amp\`{e}re laws along with the dynamics of the scalar field. However, we assess that the way we define the electromagnetic fields $\vec{E}$ and $\vec{B}$ leads to different physical interpretations of the nature of the charge and of the scalar fields. For example we could consider either a variable charge in a homogeneous vacuum or a constant charge in an inhomogeneous medium. We make the arbitrary choice of the second interpretation to describe the propagation of the GNSS signals. To be consistent with this interpretation, the electromagnetic fields emitted and received by the GNSS satellite and receiver, respectively, have to be defined as: 
\begin{displaymath}
 E^i = c\, \delta^{ij}\, F_{0j} \quad \textrm{and} \quad B^i = -\frac{1}{2}\, \varepsilon^{ijk}\, F_{jk} \, ,
\end{displaymath}
such that the Lorentz force acts on a constant charge. The constant $c$ denotes the speed of light in vacuum while $\epsilon^{ijk}$ is the totally antisymmetric tensor such that $\epsilon^{123}=1$ in the Euclidean space.


Hence, the LPI violating Lagrangian density  yields (\ref{eq:Maxwell-phi_Lag}) the following Maxwell's equations:
\begin{eqnarray}\label{eq:Maxwell_gen-B[phi]}
\frac{k_0}{c^2}\, \vec{\nabla} \cdot \left[\left(1\pm \frac{\varphi^{\mathrm{n}}}{\Lambda^{\mathrm{n}}} \right) \vec{E}\right] & = & \rho_{\rm{f}}  \\
k_0\, \vec{\nabla} \times \left[ \left(1\pm \frac{\varphi^{\mathrm{n}}}{\Lambda^{\mathrm{n}}} \right)\, \vec{B}\right] -  \frac{k_0}{c^2}\,  \partial_t \left[\left(1\pm \frac{\varphi^{\mathrm{n}}}{\Lambda^{\mathrm{n}}} \right)\, \vec{E}\right] & = & \vec{J_{\rm{f}}} 
\, . \nonumber
\end{eqnarray}
Finally the properties of the antisymmetric tensor $F_{\mu\nu}$ (Bianchi identities) complete the set of Maxwell's equations,
\begin{eqnarray}\label{eq:Maxwell-Bianchi}
\vec{\nabla} \cdot \vec{B} = 0 \, , & & \vec{\nabla} \times \vec{E} = - \partial_t \vec{B} \, ,
\end{eqnarray}
namely the Maxwell-Faraday's law and the ``no magnetic monopole'' law. As announced, we recover formally that the set of equations (\ref{eq:Maxwell_gen-B[phi]}) and (\ref{eq:Maxwell-Bianchi}) describes the propagation of electromagnetic waves in an inhomogeneous medium with specific electromagnetic properties. Indeed defining the electric displacement $\vec{D}(x)$ and the magnetizing field $\vec{H}$ respectively as: 
\begin{eqnarray}\label{def:Constitutive_B[phi]}
\vec{D} = \frac{k_0}{c^2}\, \left(1+ \frac{\varphi^{\mathrm{n}}}{\Lambda^{\mathrm{n}}} \right) \vec{E}\, ,
& \, \textrm{and} \, &
\vec{H} = k_0 \left(1+ \frac{\varphi^{\mathrm{n}}}{\Lambda^{\mathrm{n}}} \right)\, \vec{B}\, ,
\end{eqnarray}
the macroscopic Gauss' and Maxwell-Amp\`ere's laws in a medium are recovered through Euler-Lagrange equations extracted from (\ref{eq:Maxwell-phi_Lag}):
\begin{eqnarray}
\vec{\nabla} \cdot \vec{D}_{\mathrm{el}} & = & \rho_{\rm{f}}\, , \,\, \textrm{with}\,\, \rho_{\rm{f}} = \frac{J^0}{c}\, , \nonumber\\
\vec{\nabla} \times \vec{H}_{\textrm{mg}} - \partial_t \vec{D}_\textrm{el} & = & \vec{J_{\rm{f}}} \nonumber\, . 
\end{eqnarray}
We further notice that the least action principle implies that the constitutive equations (\ref{def:Constitutive_B[phi]}) are covariant which is not necessarily the case for general descriptions of relativistic electrodynamics in a moving medium (see for example \cite{McCall:2007}).

\subsection{The vacuum permittivity revisited}
Staying at the level of a first approximation, the original paper of \cite{Derevianko:2013oaa} considered a toy model of domain walls as step functions unrelated to an (effective) Lagrangian density. Although it is enough when analysing the effect on atomic clocks, this approximation scheme is too rough to cover the effects on signal propagation. In order to make the Lagrangian density (\ref{eq:Maxwell-phi_Lag}) compatible with basic domain wall solutions, we opt for the so-called Goldstone model for the dynamics of $\varphi$ with quartic $Z_2$ symmetry breaking potential $V[\varphi]$,
\begin{equation}\label{def:DW-potential}
 V[\varphi] = \frac{\lambda}{4}\, \left[ \varphi^2(x) - v^2 \right]^2 \, .
\end{equation}
The scalar field $\varphi(x)$ is assumed to have vacuum expectation values (vev) which corresponds to the minima at $\varphi=-v$ and $\varphi=v$ of this ``Mexican-hat'' shaped potential. The self-interaction constant if denoted by $\lambda$.
We make the hypothesis that the phenomenology is enough generic to be shared with more complex domain wall configurations and that the order of magnitude of experimental constraints remains basically the same. In our work, the electromagnetic waves cross the wall's core when the $Z_2$ symmetry is restored (that is, the metastable point of the quartic potential $V[\varphi]$ (\ref{def:DW-potential}) where $\varphi = 0$). Hence, our effective Lagrangian density must be explicitly $Z_2$ invariant. 
Therefore, in this paper, we restrict our analysis to a quadratic coupling with the Maxwell's sector, so that $n=1$ in (\ref{eq:Maxwell-phi_Lag}). 

In order to make the link with RF engineering and macroscopic phenomenology, we now express the interaction in the (revised) SI. The GNSS signals travel from the satellite to the receiver in an isotropic but inhomogeneous medium. Then, it makes sense to define a ``relative spacetime-dependent permittivity'' of the vacuum,
\begin{displaymath}
 \frac{\varepsilon_0(x)}{\bar{\varepsilon}_0} = 1 \pm \frac{\hbar\, c\, \left[ \varphi^2(x) - v^2 \right]}{\Lambda^2 \pm \hbar\, c\, v^2}\, ,
\end{displaymath}
where $\bar{\varepsilon}$ is the vacuum permittivity in [F/m] far from the core of the domain wall, where the $\varphi$-field is in its vev:
\begin{displaymath}
 \bar{\varepsilon}_0 = k_0\, \left( 1 \pm \frac{\hbar\, c}{\Lambda^2} v^2 \right) \, .
\end{displaymath}
We also assert that the Lorentz covariance implies that the electric polarization of the medium exactly compensates for its magnetization: 
\begin{equation}\label{eq:Relative_vacuum-permittivity_B(phi)}
 \mu_0(x) = \frac{1}{\varepsilon_0(x)\, c^2} \, .
\end{equation}
We obviously assume that most of the time, the receiver devices on Earth are far away from any domain wall. Therefore, in the former definition of the SI, the constant $\mu_0$ corresponded to the arbitrary value :
\begin{displaymath}
 \mu_{0}^{\textrm{old}} = 4\pi\, 10^{-7} \, \left[ \frac{\textrm{H}}{\textrm{m}} \right]\, .
\end{displaymath}
In the revised SI, the interpretation of the vacuum permeability $\mu_{\mathrm{0}}$ as an experimentally determined parameter remains controversial, as explained in \cite{Goldfarb:2017}. In our model, it becomes a phenomenological measurement of variations of the vacuum permeability within an interpretation of the vacuum as an inhomegeneous medium.

The Lagrangian density of our model in the SI thus reads: 
\begin{equation}\label{Lag:Phi-Soliton_Maxwell}
 \mathcal{L}^{\textrm{M+DM}} = -\frac{c^2\, \bar{\varepsilon}_0}{4}\,\, \frac{\Lambda^2 \pm \hbar\, c\, \varphi^2(x) }{ \Lambda^2 \pm \hbar\, c\, v^2 }\,  F_{\mu\nu}\, F^{\mu\nu} + \frac{1}{2} \partial_\mu \varphi\, \partial^\mu \varphi - V\left(\varphi\right) + \cdots  \, , 
\end{equation}
where the Faraday tensor is in [T]. As expected, the usual Maxwell theory in the vacuum is thus recovered if $\varphi(x)$ is frozen in one of its vev, as shown by (\ref{def:DW-potential}).  Notice that this model and all the discussion above could be easily extended to more complex domain wall configurations with several scalar fields and adequate self-interacting potentials, in particular networks of domain walls described e.g. in \cite{Sousa:2015cqa}. 

In our model, the fine structure constant $\alpha_\textrm{eff}(x)$ varies significantly only inside the wall according to the classical rescaling:
\begin{displaymath}
 \frac{\alpha_\textrm{eff}(x)}{\alpha_0} = 1 \mp \frac{\hbar\, c \, \left[\varphi^2(x)-v^2\right]}{\Lambda^2 \pm \hbar\, c\, \varphi^2(x)} \, ,
\end{displaymath}
where $\alpha_0$ is the measured value far away from the core of the domain wall. Our method of using the RF signal propagation introduces a new way of measuring the variations of $\alpha$. Indeed what is measured in that case is the relative space variation of $\alpha$ along the signal's path instead of the difference of $\alpha$ in two distinct locations. Using the analogy with optics, our results depend on the inhomogeneities of the medium along the trajectory of the RF waves, encoded in the free-space permeability.

\section{Signal propagation}
\subsection{The modified plane wave equations}
%
The macroscopic Lorentz-covariant Maxwell's equations (\ref{eq:Maxwell_gen-B[phi]}) and (\ref{eq:Maxwell-Bianchi}), describing the propagation of the GNSS signals inside a domain wall, admit modified plane-wave solutions.
We analyse the expected phenomenology of the transmission of GNSS signals in a simplified framework.
We consider a local inertial reference frame in which a satellite and a receiver are in space and at rest with the domain wall configuration. Considering the planar geometry, we assume that the wall is perpendicular to the $z$ direction so that its properties are independent of the ($x$,$y$) coordinates. The core of one domain wall is located at $z=0$ in Cartesian coordinates. Then, the scalar field is only $z$-dependent as well as the vacuum permittivity and permeability. Setting down this \ansatz along with the definition (\ref{eq:Relative_vacuum-permittivity_B(phi)}), the wave equation for the electric field far from the source reads:
\begin{displaymath}
 \nabla^2 \vec{E} - \frac{1}{c^2}\, \frac{\partial^2 \vec{E}}{\partial t^2} = - \tilde{\varepsilon}''\, E^z\, \vec{1}_z - \tilde{\varepsilon}'\, \frac{\partial \vec{E}}{\partial z}\, , 
\end{displaymath}
where $\vec{1}_z$ is the unit vector in the $z$-direction and we where have defined:
\begin{displaymath}
 \tilde{\varepsilon} (z)= \ln{\left[\frac{\varepsilon_0(z)}{\varepsilon_0}\right]} \quad \textrm{and} \quad \tilde{\varepsilon}' = \frac{\rmd \tilde{\varepsilon}}{\rmd z} \, .
\end{displaymath}
At the level of the effective material physics, by analogy with the notations of \cite{Landau:1960} and \cite{Born:1999} describing the theory of dielectric films, we consider a monochromatic time-harmonic wave with an angular frequency $\omega$ which propagates along the $xz$ plane in a medium where the permittivity only varies in the $z$-direction. In such a scheme all the electromagnetic fields are independent of the $y$-coordinate. The uniformity of the medium in the $x$-direction allows to define the electric field components as  
\begin{equation}\label{def:factorisation_E}
 E^\rmj(x,z,t) = U^\rmj(z)\, \rme^{[\textrm{i}(\kappa x-\omega t)]},
\end{equation}
where $\kappa$ denotes the $x$-component of the wave vector and so $\kappa = 0$ in the case of a normal incidence.

When the wave is linearly polarized with its electric vector normal to the plane of incidence, the wave is referred to as being ``transverse electric'' and is denoted by ``TE''. The modified wave equation for the electric field then reads:
\begin{equation}\label{eq:TE-wave_Ey}
\frac{\partial^2 E^y}{\partial z^2} + \gamma^2\, E^y = -\tilde{\varepsilon}'\, \frac{\partial E^y}{\partial z}\, , 
\end{equation}
where the constant parameter $\gamma$ defined as:  
\begin{displaymath}
 \gamma^2 = \frac{\omega^2}{c^2} - \kappa^2\, ,
\end{displaymath}
plays the role of the initial $z$-component of the wave vector before crossing the wall.

\subsection{The phase shift a the second order}

Using the factorisation (\ref{def:factorisation_E}) for the TE wave's electric field, the modified wave equation (\ref{eq:TE-wave_Ey}) reduces to:
\begin{equation}\label{eq:TE-wave-U}
 \frac{U''}{\gamma^2} + U = - \frac{\tilde{\varepsilon}'}{\gamma^2}\, U' \, , \quad \textrm{with}\, \, U' \equiv  \frac{\rmd U(z)}{\rmd z} \, ,
\end{equation}
in which we used the shorter notation $U^y(z) \equiv U(z)$. The left-hand side of (\ref{eq:TE-wave-U}) is the usual wave equation describing a plane wave propagating in the vacuum with a phase velocity $c$.
The WKB(J) method is named after Wentzel, Kramers, Brillouin and Jeffreys. It consists in expanding the complex-valued argument of an exponential function up to a given order. 
The goal is to find a solution to differential equations like the equation (\ref{eq:TE-wave-U}). 
Our approximation scheme holds provided that the successive derivatives of the relative permittivity stay small compared to the wave number, 
\begin{displaymath}
 \frac{\tilde{\varepsilon}^{\mathrm{n}}{}'}{\gamma^{\mathrm{n}}} << 1\, ,
\end{displaymath}
and that $1/\gamma$ be small compared to the characteristic lengths of the system.
Thus, the following asymptotic series expansion of the argument of an exponential function,
\begin{equation}\label{def:WKB_expansion}
 U(z) \propto \exp\left[ \gamma \sum_{\rmj = 0}^\infty \gamma^{-\rmj}\, S_\rmj(z) \right]\, ,
\end{equation}
is assumed to be a solution of (\ref{eq:TE-wave-U}) in the limit $\gamma\to\infty$. 

In our case, the imaginary part $\mathcal{S}(z)$ of the WKB(J) series expansion (\ref{def:WKB_expansion}) is given by the even orders,
\begin{displaymath}
 \rmi\, \mathcal{S}(z) = \sum_{\rmj = 0}^\infty \gamma^{-2\rmj}\, S_{2\rmj}(z)\, ,
\end{displaymath}
and is called the $\textit{optical path}$. A {\it wave front} is defined as a surface of constant phase $\mathcal{S}$.  
The {\it wave vector} is defined as being the vector normal to the wave front and thus tangent to the light ray, with $\vec{k}= \vec{\nabla} \phi$. The geometrical optics approximation consists in taking only the order 0 and so the Eikonal equation is recovered, 
\begin{displaymath}
 \left( \frac{\rmd S_0}{\rmd z} \right)^2 = -1 \, .
\end{displaymath}
This equation implies that the wave propagates in the domain wall like in the vacuum, since the phase of a plane wave with a velocity $c$ is recovered,
\begin{displaymath}
 \phi_0(x,z,t) = \kappa x + \gamma z - \omega\, t \, ,
\end{displaymath}
thus the refractive index of the medium is unitary, $n = 1$.
The domain wall has no influence on the propagation of the electromagnetic wave at this order.

At the order 2, our results reveal that the GNSS signals propagate inside the wall with a further phase shift of the carrier:
\begin{equation}\label{eq:Phase_DW}
 \phi(x,z,t) = \phi_0(x,z,t) - \gamma\left( \frac{2\tilde{\varepsilon}'+\int \left(\tilde{\varepsilon}' \right)^2 \rmd z}{8\gamma^2}\right) \, ,
\end{equation}
as far as the TE field is concerned. The second term in the right hand side of (\ref{eq:Phase_DW}) induces an extra time delay on the signal propagation measurable by ground receivers. The main observable effects in GNSS data should be:
\begin{itemize}
 \item A delay due to the bending. The ephemeris of the GNSS satellites being known quite precisely, the bending of the trajectory (longer optical path) implies a further delay which impacts the carrier phase measurement.    
 \item A delay due to the slowing down. If the phase velocity turned out to be affected, a subsequent delay should be revealed by the carrier phase measurement.
\end{itemize}
These two items share strong analogies with the assessment of the tropospheric and ionospheric effects. The domain wall can however not be reduced to a stratified medium as atmospheric effects.
The effect of interest is at the second order beyond the approximation of the geometrical optics, hence allowing curved trajectory for the rays. As there are two terms in the expression of the phase and two effects for each term, a total of four effects are described below: phase variations, transient phase shift, local bending and modification of the apparent position.  


\section{Qualitative description}
\subsection{Phase delay}
The variation of the phase velocity inside the domain wall generates a time delay $\Delta\, t$ which reads:
\begin{displaymath}
 \omega\, \Delta t = \phi(z_\textrm{r}) - \phi(z_\textrm{e})\, ,
\end{displaymath}
in comparison with the propagation in the vacuum.
The coordinates $z_{\mathrm{e}}$ and $z_{\mathrm{r}}$ are the respective positions of the emitter and the receiver in the line of sight.
At the second order, the phase delay of the TE wave between the emitter and the receiver induced by the domain wall crossing is given by (\ref{eq:Phase_DW}):
\begin{equation}\label{eq:Phase_shift}
 \Delta \phi =  \frac{1}{8\gamma}\, \left[ 2\Delta\tilde{\varepsilon}'+\int_{z_{\mathrm{e}}}^{z_{\mathrm{r}}}  \left(\tilde{\varepsilon}' \right)^2 \textrm{d} z\right] \, .
\end{equation}

\begin{figure}[h]
\centering
\includegraphics[width=0.7\linewidth]{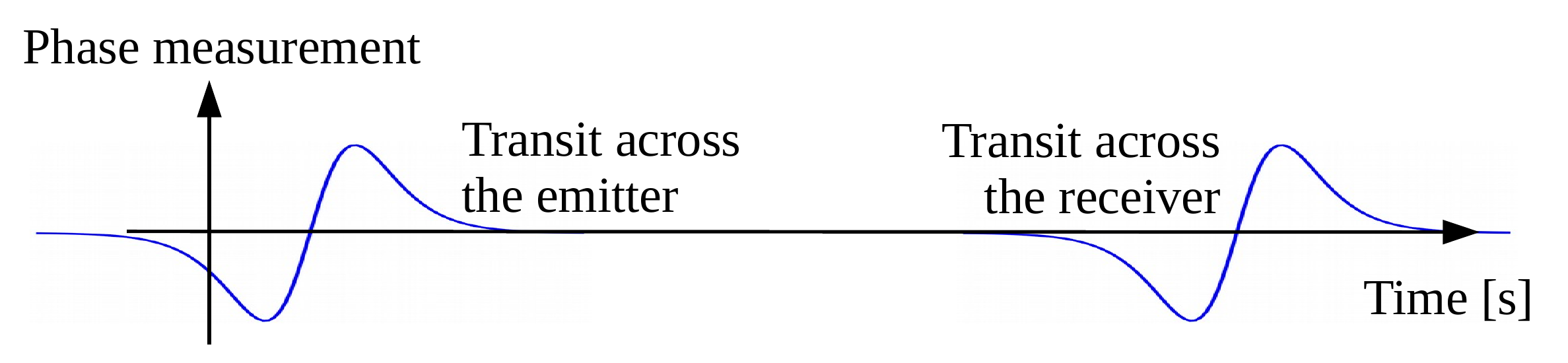}
\caption{\small{Qualitative view of the transient effect of a domain wall obtained from the first term of (\ref{eq:Phase_shift}). The phase variation is measured at the level of the GNSS receiver.\label{fig:Effect-delay_1}}}
\end{figure}

\paragraph{Phase variations} This effect on phase delay can be isolated if we consider only the first term in the right hand side of (\ref{eq:Phase_shift}). As illustrated in fig.\ref{fig:Effect-delay_1}, it only occurs during the transit of the domain wall across the emitter and the receiver. The impact on the GNSS measurement is a phase variation when the domain wall crosses the satellite, and another phase variation when the domain wall crosses the receiver. 

\begin{figure}[ht]
\centering
\includegraphics[width=0.65\linewidth]{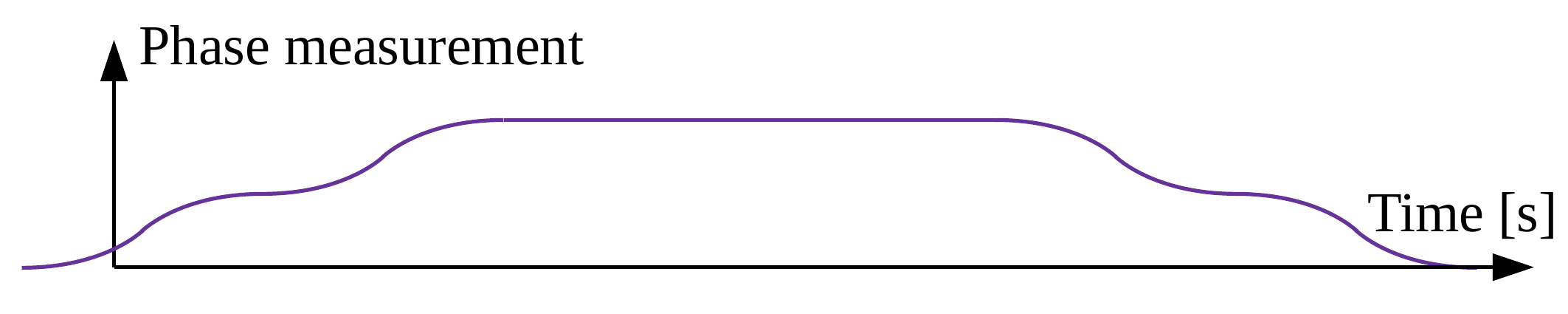}
\caption{\small{Qualitative view of the transient effect of a domain wall obtained from the second term of (\ref{eq:Phase_shift}). The phase shift is measured at the level of the GNSS receiver.\label{fig:Effect-delay_2}}}
\end{figure}

\paragraph{Transient phase shift} 
This effect is induced by the integral in the right hand side of (\ref{eq:Phase_shift}). Therefore, as illustrated in fig.\ref{fig:Effect-delay_2}, it possesses the noteworthy property to be cumulative during the transit of the domain wall between the emitter and the receiver. For a given satellite it offers the great advantage to be observed during a longer period. Its impact on the GNSS measurement is a phase shift during all the time the domain wall is located between the satellite and the receiver. 

\subsection{Bending of the trajectory}
The wave vector $\vec{k}(z)$ is defined as the vector normal to the wave front and thus tangent to the EM ray, with $\vec{k}= \vec{\nabla} \phi$. The perturbation in its $z$-component $k_z(z)$ due to the domain wall reads:
\begin{equation}\label{eq:Wave_vector}
\Delta k_z(z) = - \left(\frac{2\tilde{\varepsilon}''+ \left(\tilde{\varepsilon}' \right)^2}{8\gamma}\right) \, .
\end{equation}
As $k_z(z)$ is $z$-dependent, the trajectory of the electromagnetic wave inside the domain wall is curved. In that case, a time delay is induced by the extra distance along which the signal travels. The total length of the trajectory of the electromagnetic ``ray'' inside the defect reads:
\begin{displaymath}
 L = \int_{-\frac{d}{2}}^{\frac{d}{2}}\sqrt{1+\left(\frac{\kappa}{k_z(z)}\right)^2}\, \textrm{d}z \, ,
\end{displaymath}

\begin{figure}[h]
\centering
\includegraphics[width=0.5\linewidth]{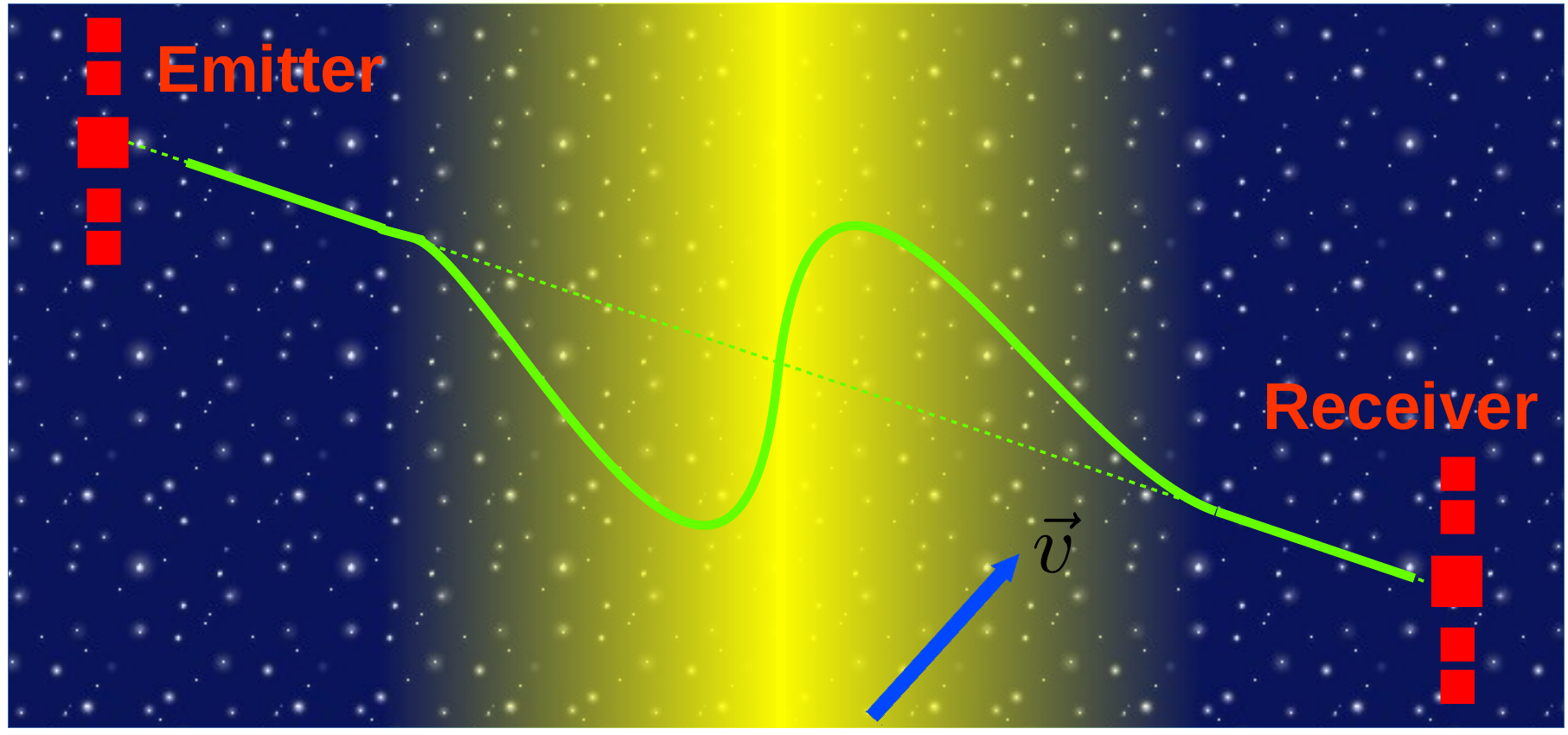}
\caption{\small{Qualitative view of the propagation of electromagnetic signals (green) across a domain wall (yellow). The associated wave vector relies on the first term in (\ref{eq:Wave_vector}).\label{fig:Effect-delay_3}}}
\end{figure}

\paragraph{Local bending} This effect, illustrated in fig.\ref{fig:Effect-delay_3}, is isolated if we consider only the first term in the right hand side of (\ref{eq:Wave_vector}).   

\begin{figure}[ht]
\centering
\includegraphics[width=0.48\linewidth]{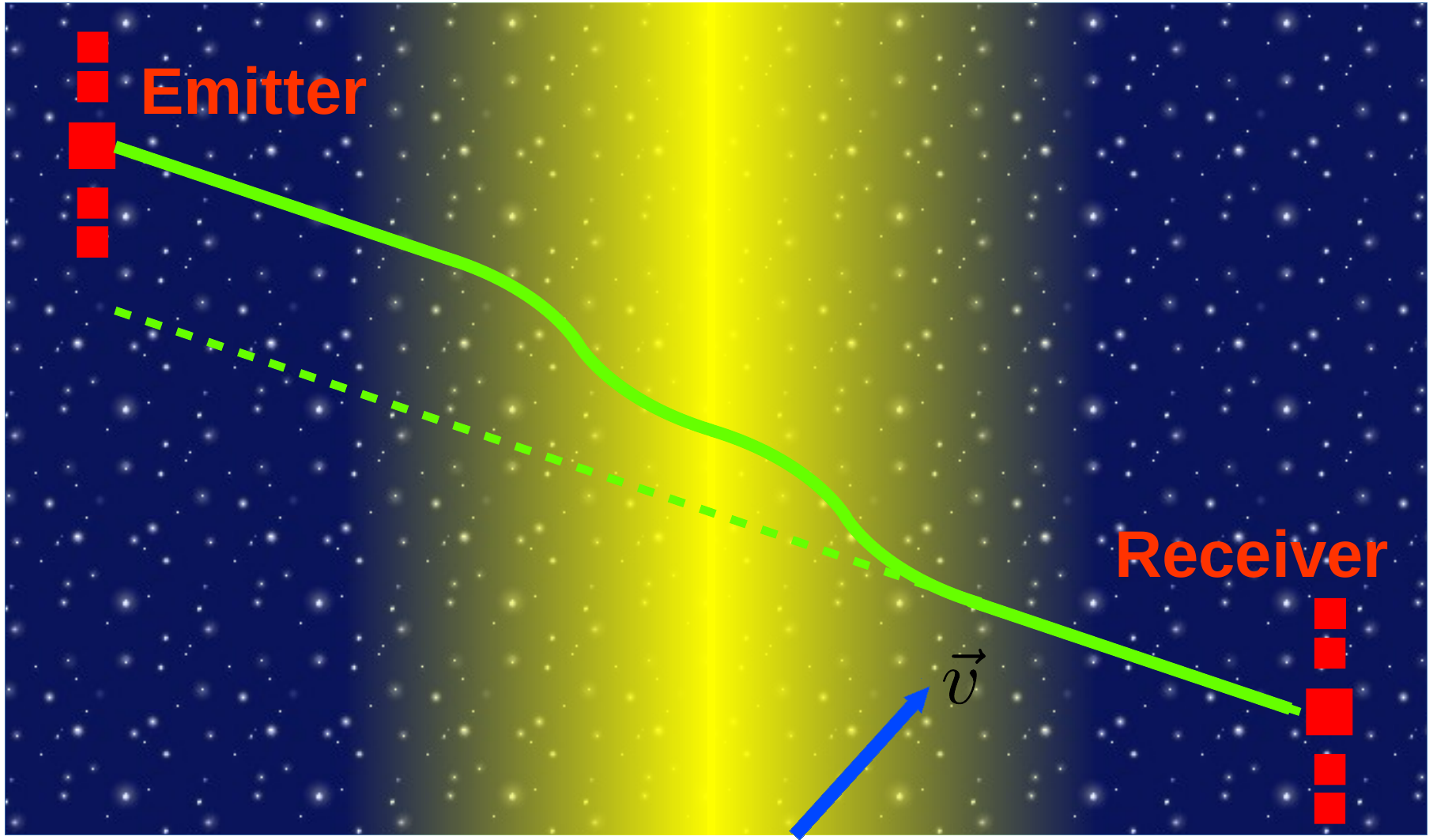}
\caption{\small{Qualitative view of the propagation of electromagnetic signals (green) across a domain wall (yellow). The associated wave vector relies on the second term in (\ref{eq:Wave_vector}).\label{fig:Effect-delay_4}}}
\end{figure}

\paragraph{Modification of the apparent position} This effect is induced by the second term in the right hand side of (\ref{eq:Wave_vector}) and is illustrated in fig.\ref{fig:Effect-delay_4}.

\section{Quantitative estimation}
In our simplified framework, an electromagnetic wave is emitted in the GNSS frequency range $\omega/2\pi$ with an incidence angle $\theta$ with the domain wall. This ideal model is in fact an approximation of the transmission of the GNSS signals from the satellite to the receiver on the ground. In this study, we consider neither the gravitational potential, nor the relative speed between the wall and the receiver or the satellite. We indeed quantified these effects as staying under the statistical noise of the GNSS measurements, far from our sensitivity threshold.  
The astrophysical parameters taken as an input are identical to \cite{Roberts:2017hla}, from the properties of the DM halo model. 
\begin{itemize}
 \item We consider a relative speed of domain wall of 300km/s.
 \item We fix the mean duration between two consecutive encounters to 7 years.
 \item We require that the energy density averaged over the domain wall network cannot exceed the measured local dark matter density of $0.3\pm0.1$ GeV cm$^{-3}$.
 \item We consider an incidence angle $\theta$ of 75$^\circ$.
\end{itemize}
We neglect the radiation energy density term in the equations describing the dynamics of the field $\varphi(x)$.
Then, throughout this paper, we deal with a basic domain wall configuration for $\varphi(x)$ of hyperbolic tangent form described e.g. by \cite{Vilenkin:2000jqa}. This configuration arises as an analytic solution of the familiar Goldstone model introduced in (\ref{Lag:Goldstone}) and (\ref{def:DW-potential}). The astrophysical constraint $\Lambda >$~10TeV advocated in \cite{Derevianko:2013oaa} does not hold for the propagation of RF signals inside domain walls\footnote{We do not provide here our demonstration based on the nature of the symmetry breaking potential.}. 

\begin{figure}[ht]
\centering
\includegraphics[width=0.5\linewidth]{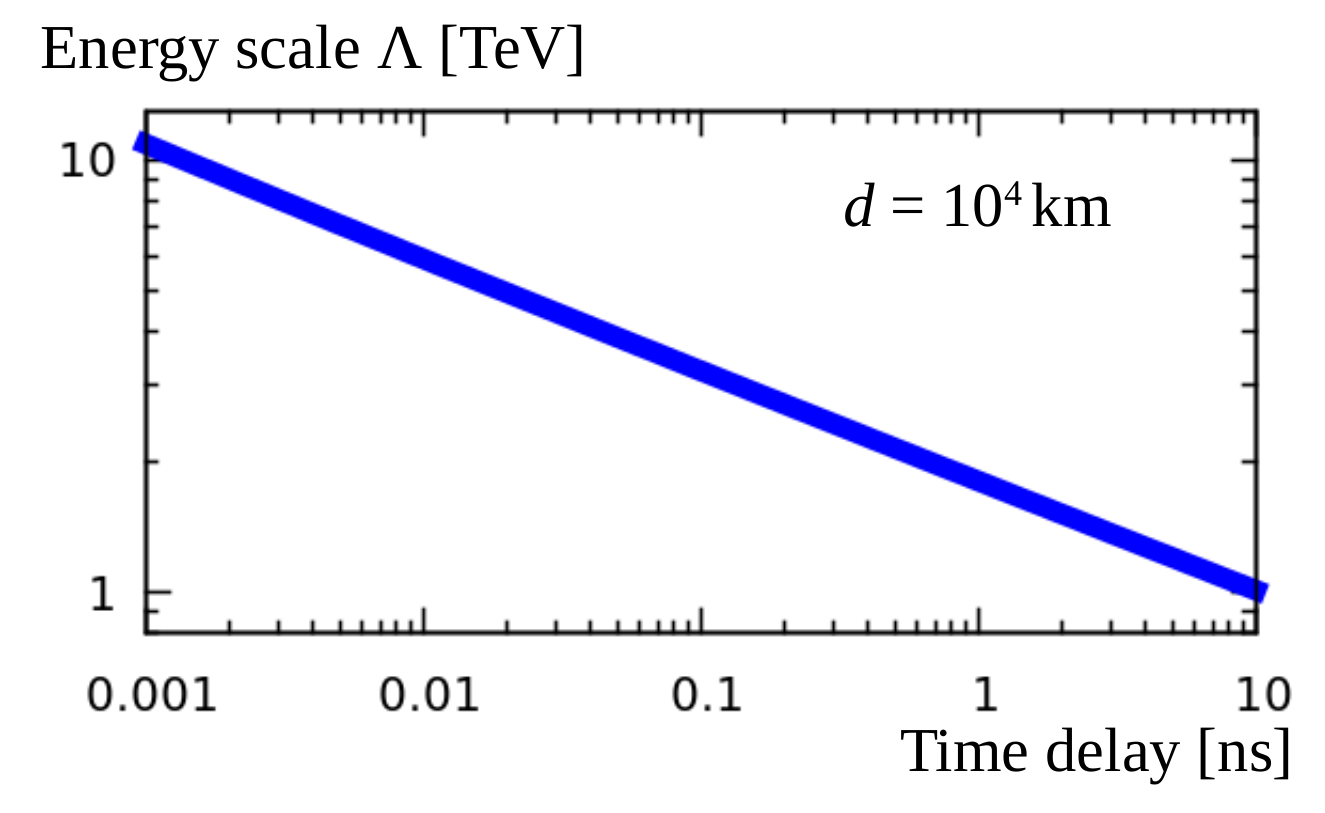}
\caption{\small{Constraints on $\Lambda$ in TeV versus the sensitivity to the time delay in ns. The excluded area is below the curve.\label{fig:Plot_Lambda-delay}}}
\end{figure}

\begin{figure}[ht]
\centering
\includegraphics[width=0.5\linewidth]{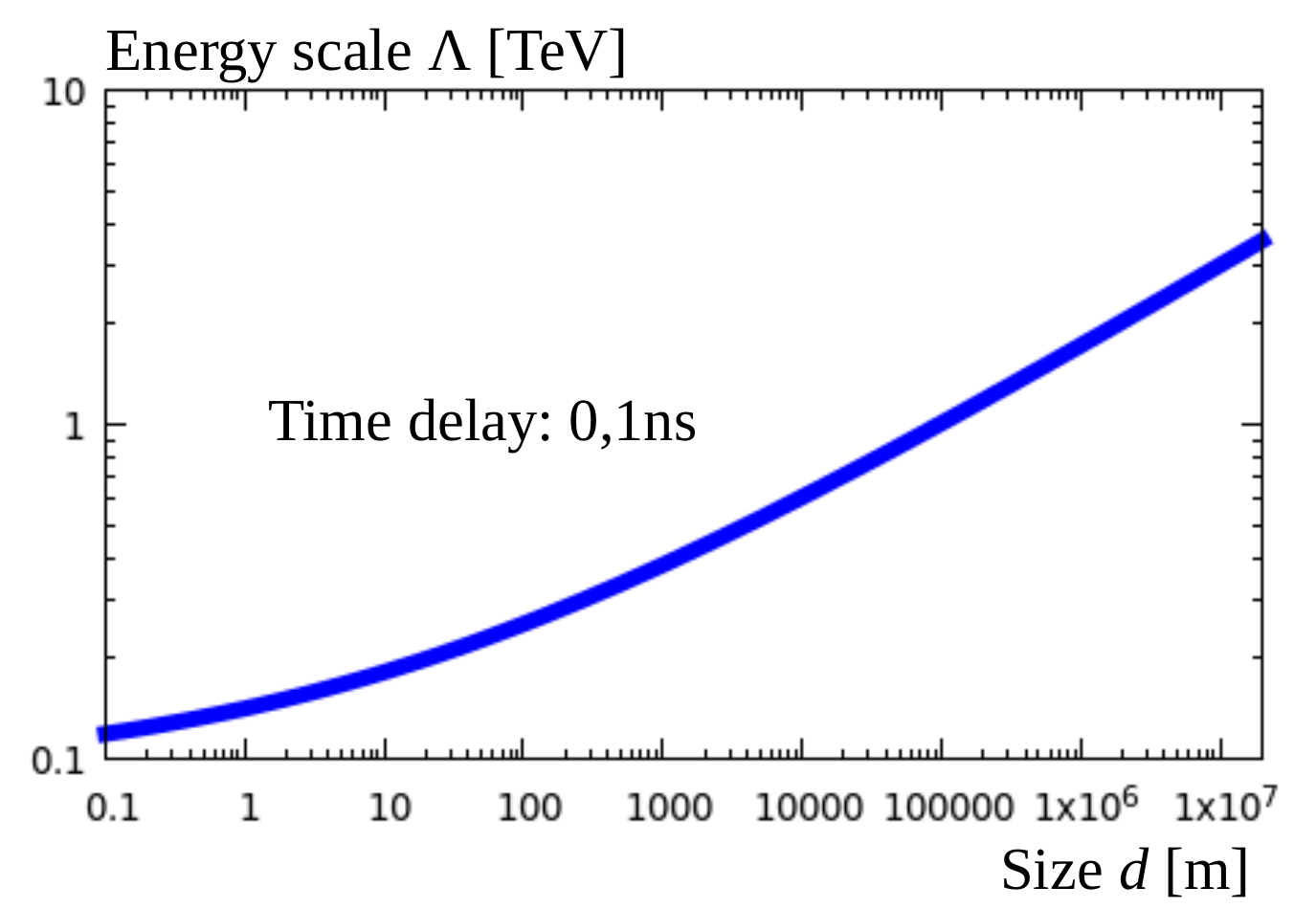}
\caption{\small{Constraints on $\Lambda$ in TeV versus the transverse size $d$ of the domain wall in m. The excluded area is below the curve.\label{fig:Plot_Lambda-d}}}
\end{figure}

Combining the four above-mentioned effects, 
we show in fig.\ref{fig:Plot_Lambda-delay} that a given sensitivity of time delay will provide constraints on the energy scale $\Lambda$. We consider here domain walls of which the thickness $d$ is in the order of the Earth diameter, that is $10^7$m. As for a recall, $\Lambda$ quantifies the interaction of dark matter with electromagnetism. The way of analyzing our results is the following: in the case of no wall signature detection during a 7-year campaign, despite a sensitivity threshold of 100ps (time delay), $\Lambda$ must be larger than 3 TeV. Our second plot in fig.\ref{fig:Plot_Lambda-d} shows that the constraints on $\Lambda$ for a given delay sensitivity also depend on the thickness $d$. It implies for example that if no wall signature is detected after a 7-year campaign despite a sensitivity of 100ps, domain walls of which the thickness is about 10m couple to EM with an energy scale greater than 200 GeV. 

The projected result in fig.\ref{fig:Plot_Lambda-delay} shows obviously that our method involving the EM links provides constraints significantly less stringent than those based on atomic clocks. However, our method is sensitive to thickness as small\footnote{This value close to the wavelength is the limit of our approximation scheme. However it is possible that even smaller values for $d$ are reachable but we did not consider this case here.} as 10cm whereas the methods based on atomic clocks know a sharp cutoff for $d<1$km, due to the servo-loop time of the GPS clocks as argued in \cite{Roberts:2017hla}. The effect of local bending is independent of the thickness $d$ of the domain wall for an approximation scheme spreading from $d$=10cm to $d$=1000km. In fact, it dominates the three other effects for thin transient domain walls ($d$ less than 10m). Therefore our technique opens the possibility to reach an unexplored region of the parameter space. Finally we also showed that the time delay is measurable for $d$ up to the solar system scale $\equiv 10^{12}$m. However, the strong constraints of \cite{Wcislo:2018ojh} and \cite{Roberts:2017hla,Roberts:2019sfo} along with the models of cosmic evolution rule out the existence of domain walls of large thickness.

\section*{Conclusion}
In this paper, we bridge the gap between a fundamental scalar interaction with electromagnetism generating a variation of $\alpha$ and a macroscopic effect on RF signals. We applied this correspondence in the case of a transient domain wall crossed by GNSS signals. At the macroscopic scale, the domain wall acts as an inhomogeneous medium with variable permittivity and permeability in which the GNSS signals propagate. This leads to an original interpretation of the ``measure'' of vacuum permeability as a test of fundamental physics. In our work, we seek for a second order effect in the phase of GNSS signals beyond the geometrical optics approximation.

While the techniques based on atomic clocks have their sensitivity to small-sized transients restricted by the transit time through the clock, the techniques involving perturbations of GNSS signals should not suffer from such restrictions. Actually, the interaction time of the signal with small-sized transients depends on the transit time between the satellite and the receiver. Thus, it opens the perspective to detect transients as small as the signal carrier’s wavelength, a size for DM transients that was not detectable previously. In the case of domain walls, the analysis of GNSS signal propagation should have a sensitivity to transverse  size ranging from a few cm to the solar system scale (10$^{12}$ m). We intend in a close future to extend this work to other types of transients like soliton solutions along with dark clusters and stars. In particular, exploring the detection of axion structure with GNSS signals turns out to be of prime interest. In this perspective, direct constraints could be brought on the abundance of such structures since the coupling to electromagnetism is inherent to the theory itself.



\end{document}